\documentclass[column,showpacs,preprintnumbers,amsmath,amssymb]{revtex4}
\preprint{}
\usepackage{graphicx}
\usepackage{dcolumn}
\usepackage{bm}
\usepackage{mathrsfs}
\begin{document}
\title{High-dimensional cryptographic quantum parameter estimation}
\author{Dong  Xie}
\email{xiedong@mail.ustc.edu.cn}
\affiliation{Faculty of Science, Guilin University of Aerospace Technology, Guilin, Guangxi, P.R. China.}

\author{Chunling Xu}
\affiliation{Faculty of Science, Guilin University of Aerospace Technology, Guilin, Guangxi, P.R. China.}
\author{Jianyong Chen}
\affiliation{Faculty of Science, Guilin University of Aerospace Technology, Guilin, Guangxi, P.R. China.}
\author{An Min Wang}
\affiliation{Department of Modern Physics , University of Science and Technology of China, Hefei, Anhui, China.}

\begin{abstract}
We investigate cryptographic quantum parameter estimation with a high-dimensional system that allows only Bob (Receiver) to access
the result and achieve optimal parameter precision from Alice (Sender). Eavesdropper (Eve) only can disturb the parameter estimation of Bob,  but she can not obtain the information of parameter. And Bob can still securely obtain a high-precision estimation of parameter by utilizing the parallel-entangled strategy and sequential strategy with a large repeat count of communication.
We analyze the security and show that the high-dimensional system can help to utilize the resource to obtain better precision than the two-dimensional system. Finally, we generalize it to the case of multi-parameter.
\end{abstract}

\maketitle

\section{Introduction}
Quantum cryptography\cite{lab1} has been the first application of quantum mechanics at the single-quantum level\cite{lab121}. Based on the laws of quantum mechanic physics, unconditional security is provided by quantum cryptograph, which performs better than classical cryptograph\cite{lab2}. And  the estimation
of physical parameters and the improvement of measurement
precision by employing quantum mechanics (quantum metrology), have attracted
considerable attention\cite{lab3,lab4,lab5,lab6,lab7,lab8,lab9}.

In ref.\cite{lab99}, V. Giovannetti et al. detail a scheme that employs entanglement and
squeezing to achieve a higher accuracy and cryptographic capabilities in position measurement, which do
not allow an eavesdropper to obtain information on the
position of Alice. And  in ref.\cite{lab999} they overcome the primary drawbacks of this scheme,
which are the difficulty of creating the requisite entanglement
and the sensitivity to loss. In ref.\cite{lab9999} they present a protocol that, using
the frequency entangled state at the output of a parametric down conversion crystal, allows one to perform quantum crypto-positioning.
In ref.\cite{lab910} G. Chiribella et al. give a simple protocol that needs no entanglement and an entangled protocol that achieves the ultimate bounds in the precision of
reference frame transmission.

 Recently, Zixin Huang et al.\cite{lab10} introduce a work for quantum cryptographic protocols specifically suited to the task of securing measurement out-comes (parameter estimation) with a two-dimensional system. In this article, we consider a new cryptographic quantum metrology with a high-dimensional probe system. Our quantum cryptographic protocol can perform better with the same number of probes than the one in\cite{lab10}. It is because that utilizing high-dimensional probes need not decoy states to detect Eve. For a single parameter, the information of parameter is randomly encoded into any two dimensions of a multi-dimensional probe. Due to the indistinguishable encoded states of probes from Alice, Eve can not obtain the detail input states.
We analyze the security and prove that Eve can not obtain the information of parameter without having been detected. Even though Eve does not worry to be detected, she still can not obtain the information of parameter due to that Alice does not tell Bob the information of prepared states with the classical communication after detecting Eve. Finally, we generalize it to cryptographic quantum multi-parameter estimation.

The rest of this article is arranged as follows. In Section II, we briefly introduce the quantum metrology of single parameter and multi-parameter, and the formula of Fisher information. In Section III, we detail the cryptographic quantum metrology protocol of a single parameter and show its security. Then, we generalize it to multi-parameter cryptographic quantum metrology protocol in section IV. A conclusion and outlook are presented in Section V.
\section{review of quantum metrology}
Quantum metrology is a fundamental subject concerning the estimation of parameters under the constraints of quantum dynamics\cite{lab6}. The famous Cram$\acute{e}$r-Rao bound\cite{lab11,lab12} offers a very good parameter estimation under the constraints of quantum physics:
\begin{eqnarray}
(\delta x)^2\geq\frac{1}{N\mathcal{F}_Q[\hat{\rho}(x)]},
\end{eqnarray}
where $N$ represents the total number of experiments.  $\mathcal{F}_Q[\hat{\rho}(x)]$ denotes quantum Fisher information (QFI), which can be generalized from classical Fisher information. The classical Fisher information is defined by
\begin{equation}
f(x)=\sum_k p_k(x)[d\ln[p_k(x)]/dx]^2,
\end{equation}
where $p_k(x)$ is the probability of obtaining the set of experimental results $k$ for the parameter value $x$. Furthermore,
the QFI is given by the maximum of the Fisher information over all measurement strategies allowed by quantum physics:
\begin{equation}
\mathcal {F}_{Q}[\hat{\rho}(x)]=\max_{\{\hat{E}_k\}}f[\hat{\rho}(x);\{\hat{E}_k\}],
\end{equation}
where positive operator-valued measure $\{\hat{E}_k\}$ represents a specific measurement device.
If the probe state is pure, $\hat{\rho}_S(x)=|\psi(x)\rangle\langle\psi(x)|$, the corresponding expression of QFI is
\begin{equation}
\mathcal {F}_{Q}[\hat{\rho}(x)]=4[\frac{d\langle\psi(x)|}{dx}\frac{d|\psi(x)\rangle}{dx}-|\frac{d\langle\psi(x)|}{dx}|\psi(x)\rangle|^2].
\end{equation}

For the classical multi-parameter Cram$\acute{e}$r-Rao bound\cite{lab5}:
\begin{equation}
 \textmd{Cov}(\widetilde{\textbf{x}})\geq F^{-1},
\end{equation}
where $\textbf{x}=\{x_1,x_2, ..., x_m\}$, $\textmd{Cov}(\widetilde{\textbf{x}})$ refers to the covariance matrix for a locally
unbiased estimator $\widetilde{\textbf{x}}(k)$, $\textmd{Cov}(\widetilde{\textbf{x}})_{jk}=\langle(\widetilde{x}_j-x_j)(\widetilde{x}_k-x_k)\rangle$ and $\langle.\rangle$ represents the average with respect to the probability distribution $p_k(\textbf{x})$.
The classic Fisher matrix for $m$ parameters as the $m\times m$ matrix with entries given by
\begin{eqnarray}
F_{jk}=\sum_i p_k(x)\left(\frac{\partial \ln[p_i(x)]}{\partial x_j}\right)\left(\frac{\partial\ln[p_i(x)]}{\partial x_k}\right).
\end{eqnarray}
\section{cryptographic quantum metrology protocol of single parameter}
The task of cryptographic quantum metrology  is that Alice sends a $d$ ($d\geq3$) dimension probe encoded with an unknown single parameter $\varphi$ to Bob, then Bob obtains the parameter by measurement. The Hilbert space of a probe can be expressed by ($|1\rangle, |2\rangle, ... ,|d\rangle$). The parameter $\varphi$ is encoded into any two levels by a unitary map $U(\varphi)$.
When Alice prepares state $\frac{\sqrt{2}}{2}(|j\rangle+|k\rangle)$ ($j<k$ and $j, k=1, 2, ..., n$), after the unitary map the encoded state is described by $\frac{\sqrt{2}}{2}(e^{-i\varphi/2}|j\rangle+e^{i\varphi/2}|k\rangle)$. For different prepared state, different unitary map is required.
After repeating $\nu$ times estimation procedure, the precision can be obtained by Eq.(4)
\begin{eqnarray}
\delta \varphi\geq 1/\sqrt{\nu}.
\end{eqnarray}
In order to improve the precision of parameter, parallel-entangled strategy and sequential strategy\cite{lab10} can improve the precision to the  Heisenberg limit.
For the parallel-entangled strategy, the encoded state of $\emph{n}$ probes is $\frac{\sqrt{2}}{2}(e^{-i n\varphi/2}|j\rangle^{\bigotimes n}+e^{i n\varphi/2}|k\rangle^{\bigotimes n})$.
For sequential strategy, after $\emph{n}$ times unitary maps the encoded state is  $\frac{\sqrt{2}}{2}(e^{-i n\varphi/2}|j\rangle+e^{i n\varphi/2}|k\rangle)$.
The corresponding precisions for two strategies are same, which is given by
\begin{eqnarray}
\delta \varphi\geq 1/(n\sqrt{\nu}).
\end{eqnarray}

Next, we consider transforming the metrology protocols into quantum cryptographically secure ones with two strategies.

\emph{Sequential strategy}-\ Firstly, for sequential strategy, cryptographical quantum metrology protocol of a single parameter can be described by the following six steps:
\begin{eqnarray}
&1.&\textmd{First step, Alice randomly prepares state}\ \frac{\sqrt{2}}{2}(|j\rangle\pm|k\rangle),\ \textmd{in which},\ j\neq k\ \textmd, j,\ k=1, 2, ..., d\ \textmd{and the operations}\nonumber \\
&&\textmd{``+" and ``-" is also chosen uniformly at random};\nonumber\\
&2.&\textmd{Second step, Alice sequentially uses \emph{n } times unitary map channel to encode the information of parameter}\ \varphi,\nonumber \\
&&\textmd{hence obtains state} \frac{\sqrt{2}}{2}(e^{-i n\varphi/2}|j\rangle\pm\nonumber
e^{i n\varphi/2}|k\rangle);\nonumber\\
&3.&\textmd{Third step, after Bob receives the encoded state from Alice (Alice determines that Bob has received the encoded}\nonumber\\
&&\textmd{state by classical communication), Alice tells Bob the measurement operators by classical communication,}\nonumber\\
&&\textmd{where the measurement operator can be described by POVM formalism (Positive Operator-Valued Measure)\cite{lab13},}\nonumber\\
&&\{E_1=\frac{1}{2}(|j\rangle+|k\rangle)(\langle j|+\langle k|),\
 E_2=\frac{1}{2}(|j\rangle-|k\rangle)(\langle j|-\langle k|),\
 E_3=1-|j\rangle\langle j|-|k\rangle\langle k|\}.\\
&4.&\textmd{ Fourth step, Bob tells Alice the measurement results. If Bob obtains the result}\ E_3,\nonumber\\
 &&\textmd{the protocol is aborted due to that the parameter has been eavesdropped by Eve.}\nonumber\\
&5.&\textmd{Fifth step, repeat the above four steps}\ \nu\ \textmd{times}.\nonumber\\
&6.&\textmd{Sixth step, Alice tells Bob the prepared states in order.}\nonumber\\
&&\textmd{ Then Bob can obtain the information parameter and estimate the precision.}\nonumber
\end{eqnarray}

Then we show that it is unconditionally secure from two cases as following: first case, Eve can not let Bob obtain the wrong information of parameter $\varphi$ without being detected; second case, Eve can not eavesdrop the information of parameter $\varphi$.

\textit{First case}-If Eve just want to let Bob obtain the wrong information of $\varphi$ without being detected, she can introduce additional $\Delta\varphi$ on the probe to bias Bob¡¯s estimation results. Due to that Eve do not know which subspace ($|j\rangle$, $|k\rangle$) is chosen each time by Alice to prepare the probe, so it is impossible to induce the same additional $\Delta\varphi$ on different encoded states.
When Eve let the encoded probes go through a fixed channel, different additional $\Delta\varphi$ is encoded into different input probes.  As a result, Bob receives the state  $\frac{\sqrt{2}}{2}(e^{-i (n\varphi/2+\Delta\varphi_{jk})/2}|j\rangle\pm e^{i( n\varphi/2+\Delta\varphi_{jk})/2}|k\rangle)$.
Then, Bob obtains different parameter value of
$\varphi$ by the measurement probability. So Eve will be detected.
Eve need to randomly introduce different additional $\Delta\varphi_{jk}$, and the expectation value of $\Delta\varphi_{jk}$ should be same $\langle\Delta\varphi_{jk}\rangle=\Delta$ for different value of (j,k). And we note that $\Delta\varphi_{jk}=-\Delta\varphi_{kj}$. It can be proved easily. For example, Eve uses a Hamiltionian $H$ to induce the additional phase. And ($|j\rangle$, $|k\rangle$) should be the eigenvectors of $H$. Otherwise, it is impossible to obtain the state $\frac{\sqrt{2}}{2}(e^{-i (n\varphi/2+\Delta\varphi_{jk})/2}|j\rangle\pm e^{i( n\varphi/2+\Delta\varphi_{jk})/2}|k\rangle)$. Then $\Delta\varphi_{jk}=(H_j-H_k)t$, where $H_j$ denotes the jth eigenvalue of $H$. So $\Delta\varphi_{jk}=-\Delta\varphi_{kj}$.
Namely the expectation value of $\langle\Delta\varphi_{jk}\rangle=-\langle\Delta\varphi_{jk}\rangle$. So $\Delta$ has to be 0.
Therefore, Bob still obtains the value of parameter $\varphi$ without a bias. Eve only reduces the precision. We consider that the probability distribution of $\Delta\varphi_{jk}$ is the Gaussian distribution $\frac{1}{\sqrt{2\pi }\delta}\exp[-\frac{(\Delta\varphi_{jk})^2}{2\delta^2}]$. The probability of result $E_1$ ($E_2$) is given by P (1-P),
\begin{eqnarray}
P=\frac{1+\cos(n \varphi) e^{-\delta^2/2}}{2}.
\end{eqnarray}
Substituting it into Eq.(6), the precision can be given by
\begin{eqnarray}
\delta\varphi\geq\frac{\sqrt{1-\cos^2 (n\varphi) e^{-\delta^2}}}{n\sqrt{\nu \sin^2 (n\varphi)e^{-\delta^2}}}.
\end{eqnarray}
Obviously, the precision of  $\delta\varphi$ is reduced. For $\varphi\neq N \pi$ ( $N=0,\pm1,\pm2,...$), the influence of Eve can be neglected by enhancing the repeat count $\nu$ and $n$.
However, for $\varphi= N \pi$, the influence of Eve can not be reduced. Then Bob gives up the result. Then they perform the cryptographical quantum metrology protocol again, but Bob measures the encoded state with a new measurement operator in the third step
 \begin{eqnarray}
 \{E_1=\frac{1}{2}(e^{-i\pi/4}|j\rangle+e^{i\pi/4}|k\rangle)(e^{-i\pi/4}\langle j|+e^{i\pi/4}\langle k|),\
 E_2=\frac{1}{2}(e^{-i\pi/4}|j\rangle-e^{i\pi/4}|k\rangle)(e^{-i\pi/4}\langle j|-e^{i\pi/4}\langle k|),\nonumber\\
 E_3=1-|j\rangle\langle j|-|k\rangle\langle k|\}.
 \end{eqnarray}
 In this way, Bob can obtain the parameter $\varphi$ with a high precision for $\varphi= N \pi$,
 \begin{eqnarray}
\delta\varphi\geq\frac{\sqrt{1- e^{-\delta^2}}}{n\sqrt{\nu e^{-\delta^2}}}.
\end{eqnarray}

\textit{Second case}-Even though Eve does not worry to be detected, she can not obtain the information of parameter $\varphi$ from the decoded states. Because, after Eve is detected, she does not know the prepared states belonging to which one of $\frac{\sqrt{2}}{2}(|j\rangle+|k\rangle)$ and $\frac{\sqrt{2}}{2}(|j\rangle-|k\rangle)$. As a result, the probability of obtaining the results $E_1$ and $E_2$ is same, so that it is impossible to obtain the information of parameter. In order to obtain the information, Eve need to conceal herself.
After Eve intercepts the encoded states, she has to send destroyed or forged states to Bob.
Eve can try to eavesdrop the information by the following two ways.

First way, Eve does not perform a measurement on the encoded state before sending a state to Bob. Eve randomly sends a state from the set $\{\frac{\sqrt{2}}{2}(|j'\rangle\pm|k'\rangle)$, in which, $j'\neq k'$ and $j', k'=1, 2, ..., d\}$. She can successfully conceal herself with the probability $(\frac{2}{d})^\nu$. For large dimension $d$ and repeat count $\nu$, Eve will be detected with the probability close to 1.
If Eve sends the state $\frac{1}{\sqrt{d}}(|1\rangle+|2\rangle+,...,+|d\rangle)$, she can conceal herself with probability $(\frac{2}{d})^\nu$. The successful probability is still close to 0. So, it is impossible to conceal herself without measurement on the encoded states in advance.

Second way, Eve performs a measurement on the encoded states and then sends a state to Bob. At this point, Eve does not know the measurement operator as shown in Eq.(9).

 If Eve chooses a projective measurement, which is given by
\begin{eqnarray}
 P_k=|k\rangle\langle k|,\
 \textmd{in which},\
 k=1, 2,...,d.
\end{eqnarray}
Then, Eve sends the projective state $|k'\rangle$ to Bob. By this measurement operator, Eve obtains nothing about the parameter $\varphi$. Bob can not detect Eve directly in the above process.
However, Bob can find that the probability of results $E_1$ and $E_2$ is same, so Bob need to give up the result. If Eve just wants to let Bob achieve the wrong information of parameter $\varphi$, she can measure a part of encoded state. This will reduce the precision of estimating parameter $\varphi$. When Eve randomly measures $m$ encoded states. The final precision of
parameter is achieved by Bob, which is given by
\begin{eqnarray}
\delta \varphi\geq\frac{\sqrt{1-(1-m/\nu)^2\cos^2 (n\varphi)}}{n(\nu-m)|\sin (n\varphi)|}.
\end{eqnarray}
For $\varphi\neq N \pi$ ( $N=0,\pm1,\pm2,...$), the influence of Eve can be neglected by enhancing the repeat count $\nu$ to be much larger than $m$. However, for $\varphi= N \pi$, the influence of Eve can be very large. So, when Bob achieves the value of parameter $\varphi= N \pi$, Bob should not trust the result. Then they perform the cryptographical quantum metrology protocol again like the above way, and Bob measures the encoded state with the measurement operator in Eq.(12).

If Eve sends a superposition state $|k'\rangle+e^{i\theta}|k''\rangle$, where $k'$ is the measurement result and $k''\neq k'$ is randomly chosen from 1 to d, and $\theta$ is a random phase factor.
Eve can conceal herself with probability $(\frac{d+1}{2d})^\nu$. For a large repeat count $\nu$, Eve can be detected with the probability of 1.

In order to obtain the information of parameter, Eve maybe use POVM
\begin{eqnarray}
 P_{jk}=\frac{1}{2d-2}(|j\rangle+|k\rangle)(\langle j|+\langle k|),\
 \textmd{in which},\
j<k=1, 2,...,d,\
P_0=1-\sum_{j<k} P_{jk}.
\end{eqnarray}
Eve can conceal herself with the probability
\begin{eqnarray}
\left\{\frac{1}{2}+\frac{3}{4(d-1)}+\left[\frac{1}{2}-\frac{3}{4(d-1)}\right]\frac{2d-1}{d(d-1)}\right\}^\nu.
\end{eqnarray}
For $\nu\gg1$, Eve must be detected. So it is very secure.
If Bob does not reveal Eve, Eve can obtain the information with the precision
\begin{eqnarray}
\delta \varphi\geq\sqrt{\frac{8[d-1-\cos^2(n\varphi/2)]\cos^2(\frac{n\varphi}{2})}{\nu n^2\sin^2(n \varphi)}}.
\end{eqnarray}
For high dimension $d$, Eve only achieve a very low precision of parameter.

Besides the above two measurement ways, Eve can also use other measurements.  However, due to the indistinguishable states (non-orthogonal states)  prepared by Alice, Eve must be detected no matter which measurement is chosen.

In one word, our cryptographical quantum metrology protocol is secure. Bob will not obtain the wrong information of parameter and the information of parameter can not be eavesdropped by Eve.

\textit{Parallel-entangled strategy}-Entangled states can also help to enhance the parameter estimation in quantum metrology\cite{lab14,lab15}. The corresponding cryptographical quantum metrology protocol of a single parameter can be modified as follows:
\begin{eqnarray}
&1.&\textmd{First step, Alice randomly prepares state of \emph{n} probes}\ \frac{\sqrt{2}}{2}(|j\rangle^{\otimes n}\pm|k\rangle^{\otimes n}), \textmd{in which}, j<k,\ j, k=1, 2, ..., d;\nonumber\\
&&\textmd{and the operations``+" and ``-" is also chosen uniformly at random};\nonumber\\
&2.&\textmd{Second step, simultaneously use \emph{n} unitary map channels to encode the information of parameter}\ \varphi,\nonumber \\
&&\textmd{hence obtain state} \frac{\sqrt{2}}{2}(e^{-i n\varphi/2}|j\rangle^{\otimes n}\pm\nonumber
e^{i n\varphi/2}|k\rangle^{\otimes n});\nonumber\\
&3.&\textmd{Third step, after Bob receives the encoded state from Alice, Alice tells Bob the measurement operators by}\nonumber\\
&&\textmd{classical communication, where the measurement operator can be described by POVM formalism,}\nonumber\\
&&\{E_1=\frac{1}{2}(|j\rangle^{\otimes n}+|k\rangle^{\otimes n})(\langle j|^{\otimes n}+\langle k|^{\otimes n}),\
 E_2=\frac{1}{2}(|j\rangle^{\otimes n}-|k\rangle^{\otimes n})(\langle j|^{\otimes n}-\langle k|^{\otimes n}),\
 E_3=1-E_1-E_2\}.\\
&4.&\textmd{ Fourth step, Bob tells Alice the measurement results. If Bob obtains the result}\ E_3,\nonumber\\
 &&\textmd{the protocol is aborted due to that the parameter is eavesdropped by Eve.}\nonumber\\
&5.&\textmd{Fifth step, repeat the above four steps}\ \nu\ \textmd{times}.\nonumber\\
&6.&\textmd{Sixth step, Alice tells Bob the prepared states in order.}\nonumber\\
&&\textmd{ Then Bob can obtain the information parameter and estimate the precision.}\nonumber
\end{eqnarray}

It is also secure like the case of sequential strategy due to the indistinguishable  prepared states. Namely, entangled states can also realize the cryptographical quantum metrology with high precision.
\section{cryptographic quantum metrology protocol of multi-parameter}
Recently, multi-parameter metrology has attracted a lot of attention\cite{lab16,lab17,lab18,lab19,lab20}. Simultaneous estimation of multi-parameter can perform better than estimating each parameter independently. We generalize the above cryptographic quantum metrology protocol of a single parameter
to the case of multi-parameter. We consider that Alice want to send $m$ parameters to Bob securely.

The cryptographic quantum metrology protocol of $m$ parameters can be summed as follows:
\begin{eqnarray}
&1.&\textmd{First step, Alice randomly prepares state}\ \frac{1}{\sqrt{m+1}}(|k_0\rangle\pm|k_1\rangle\pm|k_2\rangle\pm...\pm |k_m\rangle), \textmd{in which},\\
 &&\{k_a,\ k_b=1, 2, ..., d\}, \{a, b=0,1,2,...m\},\ d>m+1,\ \textmd{and}\ k_a\neq k_b\ \textmd{for }\ a\neq b;\nonumber\\
&2.&\textmd{Second step, sequentially use \emph{n} times unitary operators $U(\varphi_1)$ to encode the information of parameter}\ \varphi_1,\nonumber \\
&&\textmd{according to this way, encode all parameters  $\{\varphi_1,\varphi_2,...,\varphi_n\}$ on the prepared state, hence obtain the state}\nonumber\\
 &&\frac{1}{\sqrt{m+1}}(|k_0\rangle\pm e^{in\varphi_1}|k_1\rangle\pm e^{in\varphi_2}|k_2\rangle\pm...\pm
e^{i n\varphi_m}|k_m\rangle;\nonumber\\
&3.&\textmd{Third step, after Bob receives the encoded state from Alice, Alice tells Bob the measurement operators}\nonumber\\
&&\textmd{by classical communication, where the measurement operator can be described by POVM formalism}\nonumber\\
&&\{E_{1\pm}=\frac{1}{2}(\frac{1}{\sqrt{n}}|k_0\rangle\pm|k_1\rangle)(\frac{1}{\sqrt{n}}\langle k_0|\pm\langle k_1|),\
E_{2\pm}=\frac{1}{2}(\frac{1}{\sqrt{n}}|k_0\rangle\pm|k_2\rangle)(\frac{1}{\sqrt{n}}\langle k_0|\pm\langle k_2|),\nonumber\\
&&...,\
E_{m\pm}=\frac{1}{2}(\frac{1}{\sqrt{n}}|k_0\rangle\pm|k_m\rangle)(\frac{1}{\sqrt{n}}\langle k_0|\pm\langle k_m|),\
 E_{m+1}=1-E_{1+}-E_{1-}-...-E_{m+}-E_{m-}\}.\\
&4.&\textmd{ Fourth step, Bob tells Alice the measurement result. If Bob obtains the result}\ E_{m+1},\nonumber\\
 &&\textmd{the protocol is aborted due to that the parameter is eavesdropped by Eve.}\nonumber\\
&5.&\textmd{Fifth step, repeat the above four steps}\ \nu\ \textmd{times}.\nonumber\\
&6.&\textmd{Sixth step, Alice tells Bob the prepared states in order.}\nonumber\\
&&\textmd{ Then Bob can obtain the information of \emph{m} parameter by calculating the probability and estimate the precision.}\nonumber
\end{eqnarray}
When the protocol can perform without being aborted due to Eve, the probability of measurement result for $E_{j\pm}$ is given by $P_{j\pm}=\frac{1}{2m+2}[\frac{1}{m}+1\pm\frac{2}{\sqrt{m}}\cos[2n\varphi_j]]$, with $j=1,2,...,m$.
The precision of $m$ parameters can be obtained by Eq.(5) and Eq.(6)
\begin{eqnarray}
\delta\varphi_j\geq\sqrt{\frac{(m+1)^2-4m\cos^2(2n\varphi_j)}{\nu n^2\sin^2\varphi_j}}.
\end{eqnarray}
Like the case of a single parameter, Eve can not eavesdrop the information of $m$ parameters based on the undistinguished prepared states.

\section{conclusion and outlook}
The cryptographic quantum metrology of a single parameter with a high-dimensional system is studied.  The high-dimensional system can satisfy the security by preparing the indistinguishable states. Decoy-state is not necessary to detect Eve for our protocol. We analyze the security and show that it is absolutely secure for a large repeat count. And parallel-entangled strategy and sequential strategy can be utilized to improve the parameter precision. We also utilize  the techniques of multi-parameter quantum metrology and quantum cryptography to  obtain the cryptographic quantum metrology protocol of multi-parameter.

In this article, we only consider the unitary parameters. Cryptographic quantum metrology protocol of the non-unitary parameters\cite{lab21,lab22}  will worth to be the further exploration.
And a lossy channel and imperfect measurement in cryptographic quantum metrology also will be researched.

\section*{Acknowledgement}
 This research was supported by the National
Natural Science Foundation of China under Grant No. 11747008, Guangxi Natural Science Foundation 2016GXNSFBA380227 and Guangxi Base Promotion Project of Young and Middle-aged Teachers (NO.2017KY0857).

\end{document}